\documentclass[preprint,12pt,3p,twocolumn]{elsarticle}

\usepackage{graphicx}

\usepackage{amssymb}
\usepackage{amsmath}
\usepackage[caption=false]{subfig}
\usepackage{url}

\biboptions{sort&compress}

\journal{Materials Today Physics}

\begin{document}

\newcommand{\kin}{\kappa_{\text{in}}}
\newcommand{\kout}{\kappa_{\text{out}}}
\newcommand{\pcm}{cm$^{-3}$}
\newcommand{\cmc}{cm$^{3}$}

\begin{frontmatter}

\title{Phonon thermal transport in 2H, 4H and 6H silicon carbide from first principles \tnoteref{t1}}

\tnotetext[t1]{\copyright 2017. This manuscript version is made available under the CC-BY-NC-ND 4.0 license \url{http://creativecommons.org/licenses/by-nc-nd/4.0/}}

\author[label1]{Nakib Haider Protik\corref{cor1}}
\address[label1]{Boston College, Chestnut Hill, Massachusetts 02467, USA}
\cortext[cor1]{Corresponding author}
\ead{nakib.protik@bc.edu}

\author[label2]{Ankita Katre}
\address[label2]{LITEN, CEA-Grenoble, 17 rue des Martyrs, 38054 Grenoble Cedex 9, France}
\ead{ankitamkatre@gmail.com}

\author[label3]{Lucas Lindsay}
\address[label3]{Oak Ridge National laboratory, 1 Bethel Valley Rd, Oak Ridge, Tennessee 37830, USA}
\ead{lindsaylr@ornl.gov}

\author[label4]{Jes\'{u}s Carrete}
\address[label4]{Institute of Materials Chemistry, TU Wien, A-1060 Vienna, Austria}
\ead{jesus.carrete.montana@tuwien.ac.at}

\author[label2]{Natalio Mingo}
\ead{natalio.mingo@cea.fr}

\author[label1]{David Broido}
\ead{broido@bc.edu}

\begin{abstract}
Silicon carbide (SiC) is a wide band gap semiconductor with a variety of industrial applications. Among its many useful properties is its high thermal conductivity, which makes it advantageous for thermal management applications. In this paper we present \textit{ab initio} calculations of the in-plane and cross-plane thermal conductivities, $\kin$ and $\kout$, of three common hexagonal polytypes of SiC: 2H, 4H and 6H. The phonon Boltzmann transport equation is solved iteratively using as input interatomic force constants determined from density functional theory. Both $\kin$ and $\kout$ decrease with increasing $n$ in $n$H SiC because of additional low-lying optic phonon branches. These optic branches are characterized by low phonon group velocities, and they increase the phase space for phonon-phonon scattering of acoustic modes. Also, for all $n$, $\kin$ is found to be larger than $\kout$ in the temperature range considered. At electron concentrations present in experimental samples, scattering of phonons by electrons is shown to be negligible except well below room temperature where it can lead to a significant reduction of the lattice thermal conductivity. This work highlights the power of \textit{ab initio} approaches in giving quantitative, predictive descriptions of thermal transport in materials. It helps explain the qualitative disagreement that exists among different sets of measured thermal conductivity data and provides information of the relative quality of samples from which measured data was obtained.
\end{abstract}


\end{frontmatter}


\section{Introduction} \label{intro}
The hexagonal polytypes of SiC ($n$H SiC with $n$ = 2,4,6) are large, indirect band gap semiconductors. Because of their nearly three times larger band gap compared to Si, they can maintain operational performance better than Si under harsh conditions such as high temperature, high electric field intensity and frequency, high power and strong radiation. Of particular interest to us is their large lattice thermal conductivities, $\kappa$, which make them good candidates for thermal management applications. While the thermal properties of the 4H and 6H phases have been studied, there is no consensus among the various published measurements of $\kappa$ \citep{cree,slack1964thermal,morelli1993eph,burgemeister1979thermal,nilsson1997determination,wei2013thermal,morelli1993conf}. For example, in Ref. \citep{morelli1993conf} the lattice thermal conductivity parallel to the hexagonal atomic planes, $\kin$, for the 6H phase is found to be higher than that for the 4H phase. In contrast, measured data for the 4H phase from Ref. \citep{cree} gives notably higher values than those in Ref. \citep{morelli1993conf} suggesting 4H SiC has higher thermal conductivity than that of 6H SiC. In addition, even for the same $n$ in $n$H SiC there is wide variation in $\kappa$ values among the various measurements. Finally, phonon-electron scattering has been suggested to play a major role in lowering $\kappa$ \citep{morelli1993eph,wei2013thermal}. However, no rigorous calculations have been performed to assess this. In this paper we address these issues by carrying out \textit{ab initio} calculations of $\kin$ and $\kout$ in 2H, 4H and 6H SiC. We specifically answer two questions: 1) what are the trends in $\kin$ and $\kout$ among the three considered hexagonal polytypes of SiC? And, 2) how strongly does phonon-electron scattering affect $\kappa$? We find that $\kin$ is larger than $\kout$ over a wide temperature range and that $\kin$ and $\kout$ decrease with increasing $n$ in $n$H SiC.  We also find that phonon-electron scattering has a negligible effect on $\kappa$ around and above room temperature, but that it can cause significant suppression of $\kappa$ below around $100$ K.  

\section{Phonon thermal transport} \label{transport}
Phonon transport is the dominant mechanism of heat conduction in semiconductors and insulators. This is particularly true in materials such as SiC where the stiff bonding, light constituent atoms and relatively low electron concentrations make the lattice contribution to thermal conductivity much larger than that from the charge carriers. Phonon scattering due to anharmonicity (phonon-phonon scattering), structural defects, isotopes, crystal boundaries, dopants, etc. limits the thermal conductivity. An applied temperature gradient creates a non-equilibrium phonon distribution causing phonons to move diffusively through the crystal. In steady state phonon thermal transport can be described by the Peierls-Boltzmann transport equation (PBE)
\begin{equation}\label{eq:bte}
    \mathbf{v}_{\lambda}\cdot\mathbf{\nabla} T \dfrac{\partial n_{\lambda}}{\partial T} = \left(\dfrac{\partial n_{\lambda}}{\partial t}\right)_{\text{collisions}},
\end{equation}
where $\lambda \equiv (\mathbf{q},s)$ is the phonon mode with wavevector $\mathbf{q}$ and polarization $s$, $\mathbf{v}_{\lambda}$ is the phonon group velocity, $\mathbf{\nabla} T$ is the applied temperature gradient, and $n_{\lambda}$ is the non-equilibrium phonon distribution function.

For a small temperature gradient
\begin{equation}\label{eq:distfunc}
  n_{\lambda} = n^{0}_{\lambda} + \left(-\partial n^{0}_{\lambda}/\partial T\right)\mathbf{F}_{\lambda}\cdot \mathbf{\nabla} T, 
\end{equation}
 where $n^{0}_{\lambda}$ is the equilibrium phonon (Bose-Einstein) distribution. Then linearization of Eq. \ref{eq:bte} in the assumed small $\mathbf{\nabla} T$ gives \citep{wu2014,wu2012-1}
\begin{equation}\label{eq:ibte}
    \mathbf{F}_{\lambda} = \tau^{0}_{\lambda}\left(\mathbf{v}_{\lambda} + \mathbf{\Delta}_{\lambda}\right),
\end{equation}
 where $\tau^{0}_{\lambda}$ is the total phonon mode relaxation time taking into account effects of all scatterers in the system, and $\mathbf{\Delta}_{\lambda}$ is a linear function of $\mathbf{F}_{\lambda}$ given by Eq. \ref{eq:delta} in the appendix. Eq. \ref{eq:ibte} is solved iteratively with the initial choice of $\mathbf{\Delta}_{\lambda} = 0$ and by using the Matthiessen's rule: $\left(\tau^{0}_{\lambda}\right)^{-1} = \sum_{i}\left(\tau^{i}_{\lambda}\right)^{-1}$ where $i$ denotes a type of phonon scatterer. This initial guess is equivalent to working in the relaxation time approximation (RTA) where both momentum conserving normal and momentum non-conserving umklapp processes are taken to be directly thermally resistive. In the iterative scheme the normal processes redistribute the phonon population while the umklapp processes remain directly thermally resistive. As a result, the iterative scheme usually produces higher thermal conductivity than predicted by the RTA \citep{omini1995iterative}.

Near room temperature, three-phonon scattering is the main mechanism that limits $\kappa$ \citep{ziman1960electrons}. The three-phonon scattering rates are given by Eqs. \ref{eq:phphplus} and \ref{eq:phphminus}. Phonons are also scattered by mass disorder from natural isotope mixes on Si and C atoms and from substitutional defects. The phonon-isotope scattering rates are treated using a mass-variance model \citep{tamura1983isotope} given by Eq. \ref{eq:phiso}. Substitutional defects, where they are considered, are also treated using the mass-variance model.

We consider phonon-electron scattering processes where phonons are either emitted or absorbed by electrons. This leads to an additional collision term on the right hand side with Eq. \ref{eq:bte} of the following form:
\begin{align*}
    &\left(\dfrac{\partial n_{\lambda}}{\partial t}\right)_{\text{EPI}} = \dfrac{4\pi}{\hbar}\sum_{mn}\sum_{\mathbf{k}} |g^{smn}(\mathbf{k},\mathbf{q})|^{2} \nonumber \\
    &\times \bigg\{ -f^{n}_{\mathbf{k}}(1 - f^{m}_{\mathbf{k}+\mathbf{q}})n_{\lambda} \nonumber \\ 
    &+ f^{m}_{\mathbf{k}+\mathbf{q}}(1 - f^{n}_{\mathbf{k}})(1 + n_{\lambda}) \bigg\} \delta(\epsilon^{m}_{\mathbf{k}+\mathbf{q}} - \epsilon^{n}_{\mathbf{k}} - \hbar\omega_{\lambda}),
\end{align*}
where $m, n$ are electronic band indices, $g$ is the electron-phonon interaction (EPI) matrix element, $f$ is the electronic distribution function, $\epsilon$ is the electron energy, $(2\pi)^{-1}\omega$ is the phonon frequency and $\mathbf{v}$ is the electron group velocity. Expanding the phonon distribution function as done in Eq. \ref{eq:distfunc} while taking the electronic distribution to be in equilibrium, the phonon scattering rates due to the EPI are found in the RTA to be:
\begin{align}\label{eq:epi}
    &\left(\tau^{\text{EPI}}_{\lambda}\right)^{-1} = \dfrac{4\pi}{\hbar}\sum_{mn}\sum_{\mathbf{k}} |g^{smn}(\mathbf{k},\mathbf{q})|^{2}\nonumber \\
    &\times \bigg\{ f^{n}_{\mathbf{k}} - f^{m}_{\mathbf{k}+\mathbf{q}} \bigg\}
    \delta(\epsilon^{m}_{\mathbf{k}+\mathbf{q}} - \epsilon^{n}_{\mathbf{k}} - \hbar\omega_{\lambda}),
\end{align}
where the $f \equiv f^{0}$ is now the equilibrium electron (Fermi-Dirac) distribution.

The lattice thermal conductivity tensor elements are calculated using
\begin{equation}\label{eq:kappa}
    \kappa^{\alpha\beta} = \dfrac{k_{\text{B}}}{NV}\sum_{\lambda}\left(\dfrac{\hbar \omega_{\lambda}}{\text{k}_{B}T}\right)^{2}n^{0}_{\lambda}(n^{0}_{\lambda} + 1)v^{\alpha}_{\lambda}F^{\beta}_{\lambda},
\end{equation}
where $\alpha, \beta$ are Cartesian directions, $k_{\text{B}}$ is the Boltzmann constant, $N$ is the total number of $q$-points in the Brillouin zone and $V$ is the primitive cell volume.

\section{Computational methods} \label{compute}
Density functional theory (DFT) and density functional perturbation theory (DFPT) calculations are performed using the \texttt{Quantum ESPRESSO v. 5.4.0} suite \citep{giannozzi2009quantum}. We use a norm-conserving pseudopotential. The local density approximation with the Perdew-Zunger parametrization is chosen for the exchange-correlation functional. A variable-cell structural optimization is carried out on $8\times8\times6$, $8\times8\times4$ and $8\times8\times2$ $k$-meshes for the 2H, 4H and 6H structures, respectively, to find the relaxed lattice constants and atomic positions. We use DFPT to calculate the harmonic force constants and phonon dispersions. For the phonon calculations a $6\times6\times4$ $q$-mesh is used for the 2H phase, and  $6\times6\times2$ for the 4H and 6H phases. To calculate the third order anharmonic force constants we employ a finite displacement supercell method. To generate the symmetry-allowed minimal set of two-atom-displaced supercells, extract the anharmonic force constants following $\Gamma$-point DFT calculations on said supercells, and symmetrize the force constants tensor we use the \texttt{thirdorder.py} code \citep{wu2014,wu2012-2}. We use a $3\times3\times3$ supercell, corresponding to $108$, $ 216$ and $324$ atoms, respectively, for the 2H, 4H and 6H phases. A converged cut-off distance for the third-order force constants of $3.3$ \AA  $ $ is used for each.

The \texttt{ShengBTE} software \citep{wu2014} is used to calculate the $\kappa$ tensor using a locally adaptive Gaussian broadening method \citep{wu2012-1} to approximate the energy conserving delta functions appearing in the scattering rates expressions (see Eqs. \ref{eq:gauss} and \ref{eq:adaptive}). To keep the $q$-mesh spacing uniform we use $N_{x(y)}/N_{z} \approx b_{x(y)}/b_{z}$ where $N$ is the number of sampled $q$-points, $b$ is the reciprocal lattice vector component along a given Cartesian direction, and $x(y)$ and $z$ denote Cartesian directions $\hat{x} (\hat{y})$ and $\hat{z}$. Given that in $n$H SiC $b_{z} < b_{x(y)}$ and in view of Eq. \ref{eq:adaptive}, it is important to keep the $q$-mesh spacing uniform in order to prevent the adaptive Gaussian scheme from selectively setting smaller broadening for the $(0,0,q_{z})$ modes. This causes the lowest frequency phonons to have spuriously large lifetimes. Converged $q$-meshes $25\times25\times15$, $23\times23\times7$ and $25\times25\times5$ were used for the $\kappa$ calculations of 2H, 4H and 6H SiC, respectively.

The phonon-electron scattering rates, Eq. \ref{eq:epi}, are obtained by calculating the imaginary part of the phonon self-energy using  the \texttt{EPW v. 4} code \citep{ponce2016epw,giustino2007electron}. For the desired grid of phonon modes, the \texttt{EPW} code first calculates band energies and electron-phonon matrix elements \textit{ab initio} on a coarse grid of electron wave vectors.  Then, it uses maximally localized Wannier functions to interpolate the EPI matrix elements and band energies to a fine $k$-mesh. The original code uses a fixed Gaussian broadening to approximate delta functions; this requires convergence testing in tandem with that for the $k$-mesh density. We have modified the original code to employ the analytic tetrahedron method \citep{lambin1984computation} which requires convergence testing only in the $k$-mesh density. A fine $90\times90\times60$ $k$-mesh is used to calculate the phonon-electron scattering rates for 2H SiC which are then combined with the other phonon scattering rates at the RTA level using Matthiessen's rule. We checked that the calculated $\kappa$ is converged with respect to the $k$-mesh density for the temperature range considered. Calculations for 4H and 6H SiC proved to be too computationally expensive.

\section{Results and discussion} \label{results}

The results from structural optimization calculations are presented in Table \ref{tab:opt}. We find excellent agreement between calculated and measured \citep{schulz1979structure,stockmeier2009lattice} values of the in- and cross-plane lattice constants, $a$ and $c$. For the 2H SiC internal parameter we find $u(\text{C}) = 0.3754 (0.375)$. For 4H SiC the internal parameters are $u(\text{C}) = 0.1874 (0.1875)$ and $v(\text{Si}) = 0.25 (0.25)$. And for 6H SiC: $u(\text{C}) = 0.1254 (0.125)$, $v(\text{Si}) = 0.1668 (0.1667)$, $v(\text{C}) = 0.2917 (0.2917)$, $w(\text{Si}) = 0.333 (0.333)$ and $w(\text{C}) = 0.4582 (0.4583)$. The numbers in the parentheses are the ideal values for these parameters \citep{kackell1994electronic}.

\begin{table}
    \centering
    \begin{tabular}{c|c|c}
        phase &  $a$, $c$ (\AA) & $a$, $c$ (\AA) \\
             &  theory & experiment  \\
        \hline
        2H &  $3.07, 5.04$ & $3.08, 5.05$\\
        4H &  $3.07, 10.05$ & $3.08, 10.08$\\
        6H &  $3.07, 15.08$ & $3.08, 15.12$\\
    \end{tabular}
    \caption{In-plane ($a$) and cross-plane ($c$) lattice parameters of 2H, 4H and 6H SiC. Experimental results for $a$, $c$ for the 2H phase are taken from Ref. \citep{schulz1979structure} and for 4H and 6H from Ref. \citep{stockmeier2009lattice}.}
    \label{tab:opt}
\end{table}

\begin{figure}
    \includegraphics[width=1.0\linewidth]{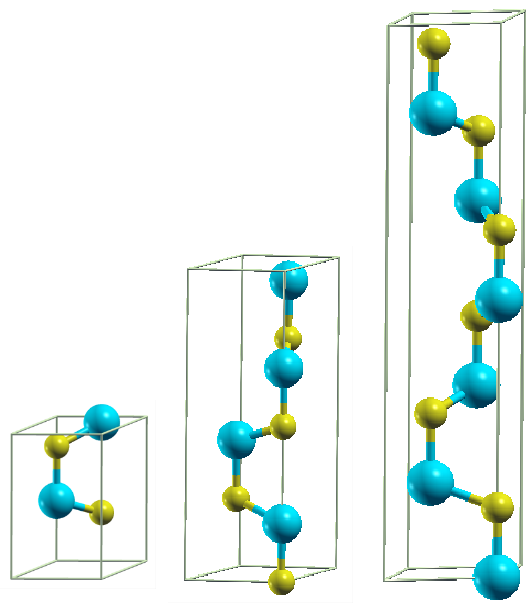}
    \caption{From left to right: 2H, 4H and 6H SiC primitive cells. Blue represents Si and yellow, C.}
    \label{fig:pcells}
\end{figure}

\begin{figure}
    \subfloat[]{%
      \includegraphics[width=1.0\linewidth]{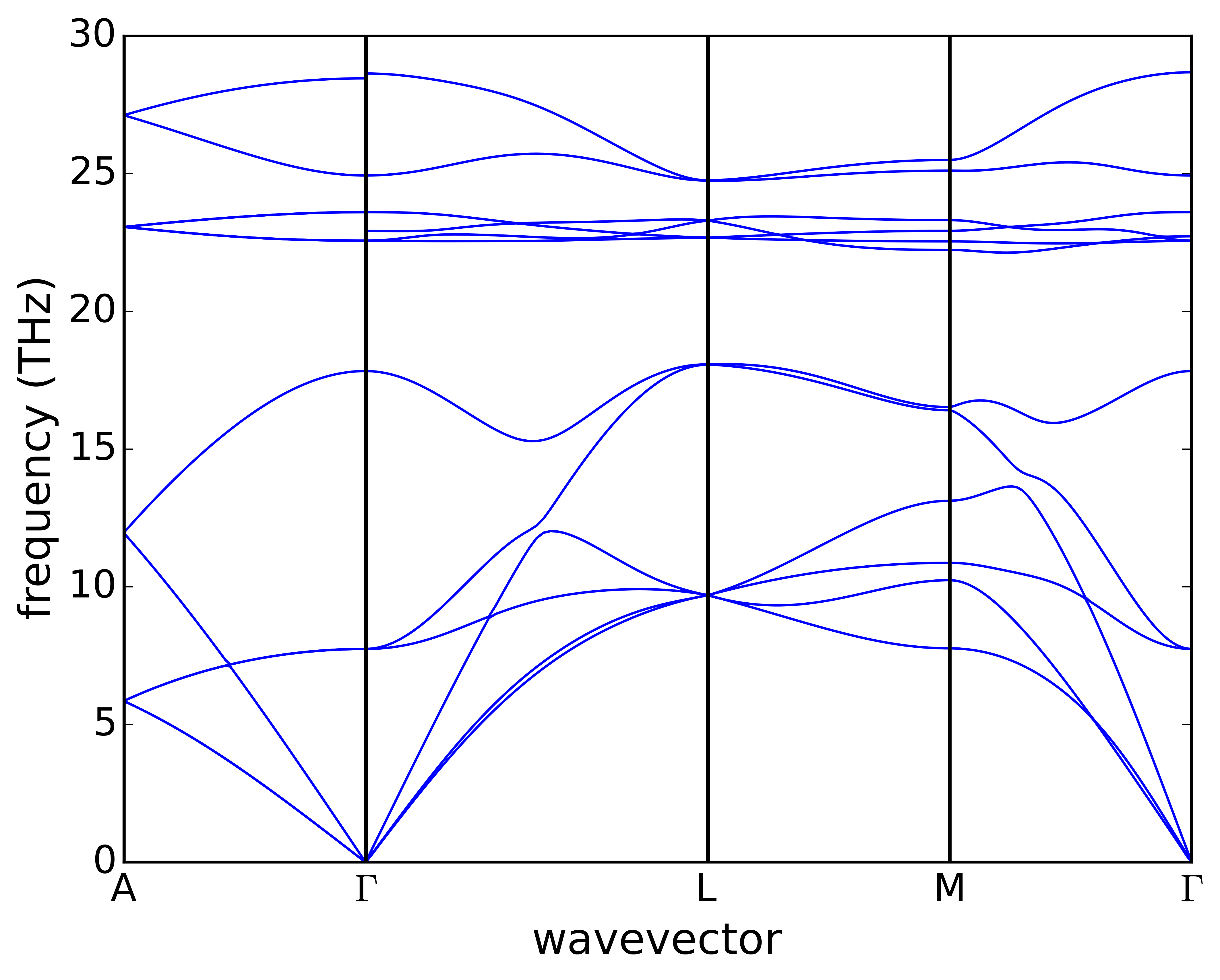}
    }\\
    \subfloat[]{%
      \includegraphics[width=1.0\linewidth]{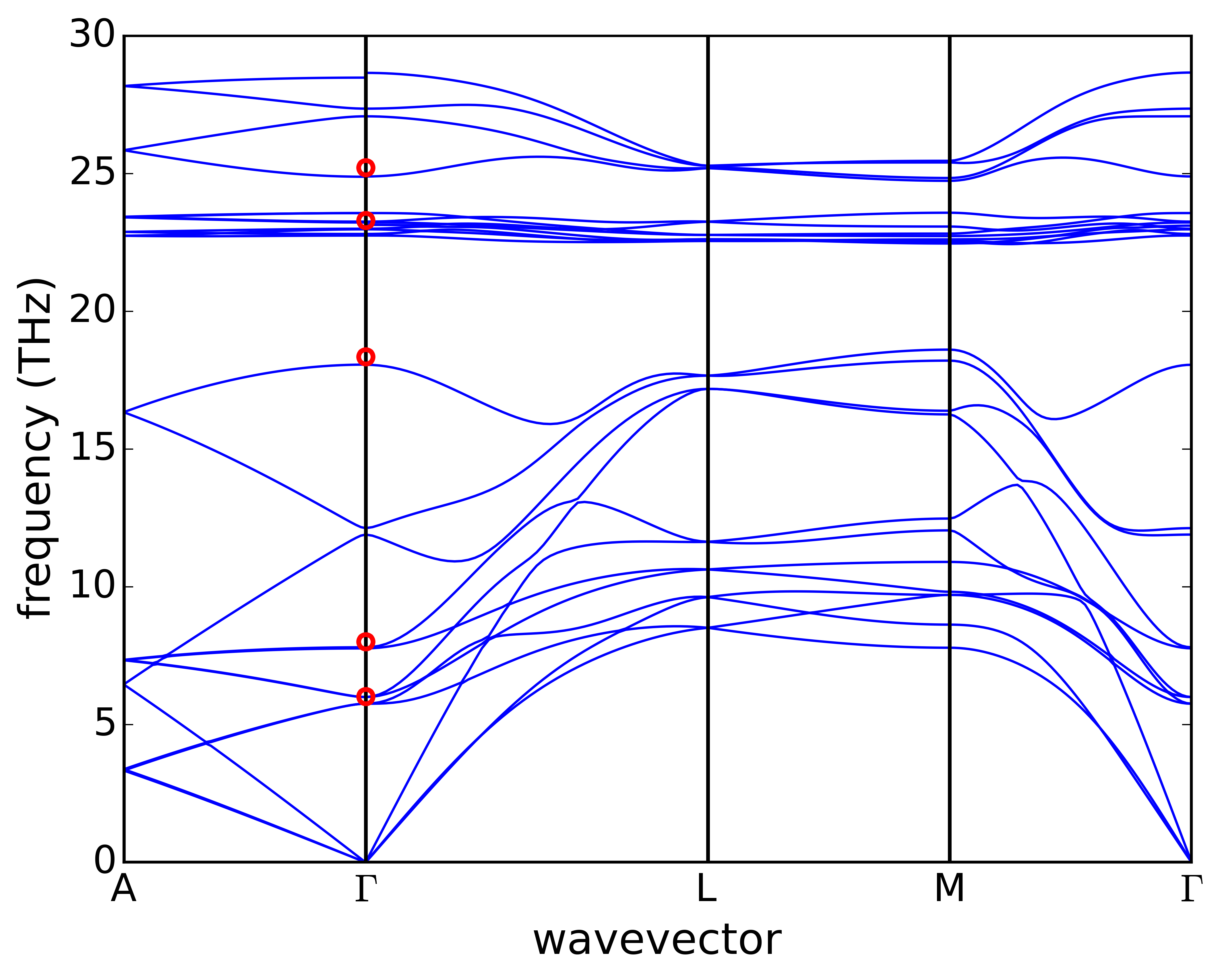}
    }\\
    \subfloat[]{%
      \includegraphics[width=1.0\linewidth]{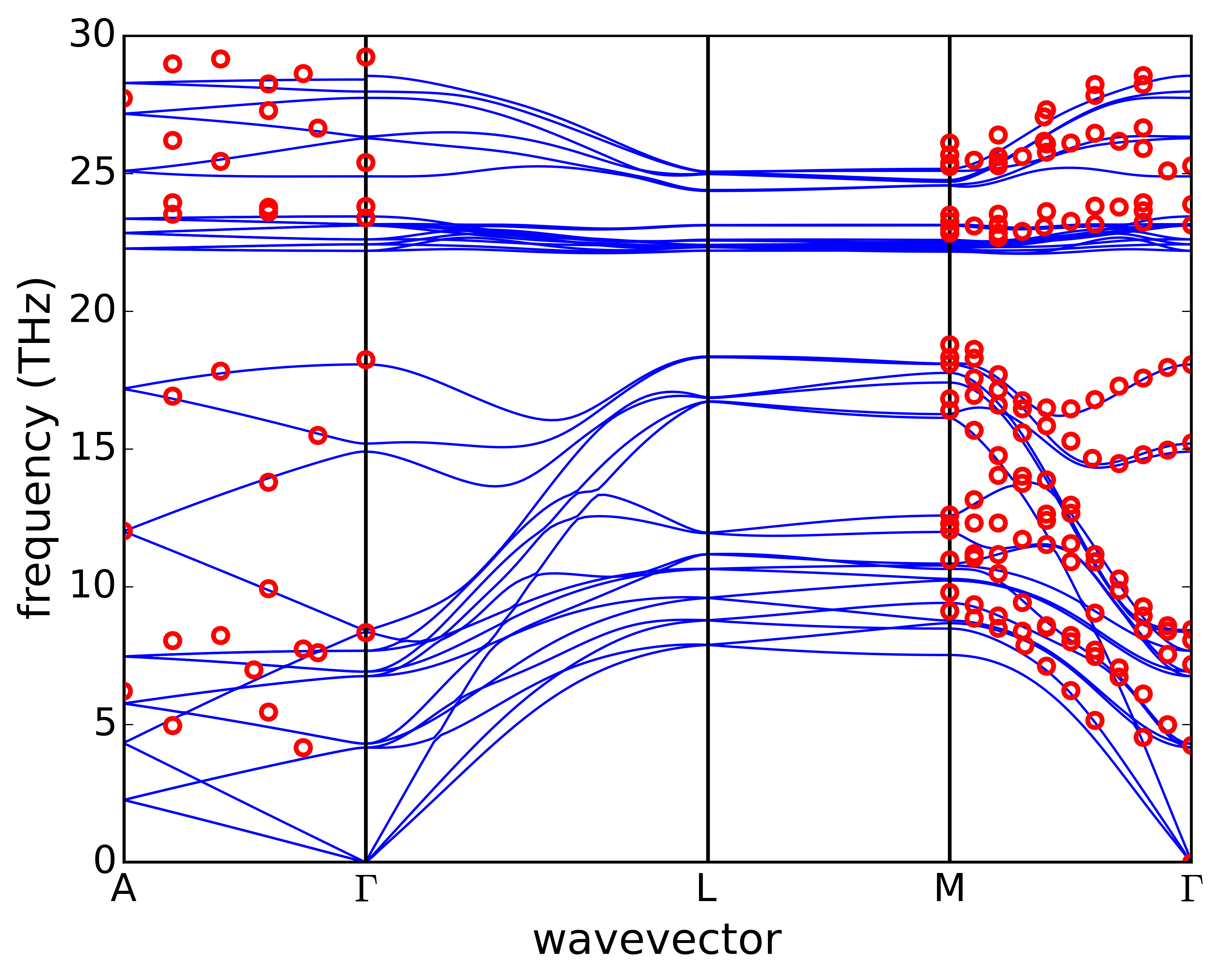}
    }
    
    \caption{Phonon dispersions of (a) 2H, (b) 4H and (c) 6H SiC. Blue curves are calculated results. Red circles are from experimental measurements for 4H \citep{feldman1968phonon} and 6H \citep{nowak2001crystal} phases, respectively.}
    \label{fig:disp}
\end{figure}

2H, 4H and 6H SiC have $4$, $8$, and $12$ atoms per primitive unit cell, respectively, with cells becoming more elongated along the cross-plane direction of $n$H SiC with increasing n as shown in Fig. \ref{fig:pcells}. Calculated phonon dispersions for 2H, 4H and 6H SiC are shown in Fig. \ref{fig:disp} along with available experimental measurements for 4H and 6H SiC \citep{feldman1968phonon,nowak2001crystal,2hdisp}. The agreement between calculation and measurement is excellent.

\begin{figure}
    \centering
    \includegraphics[width=1.0\linewidth]{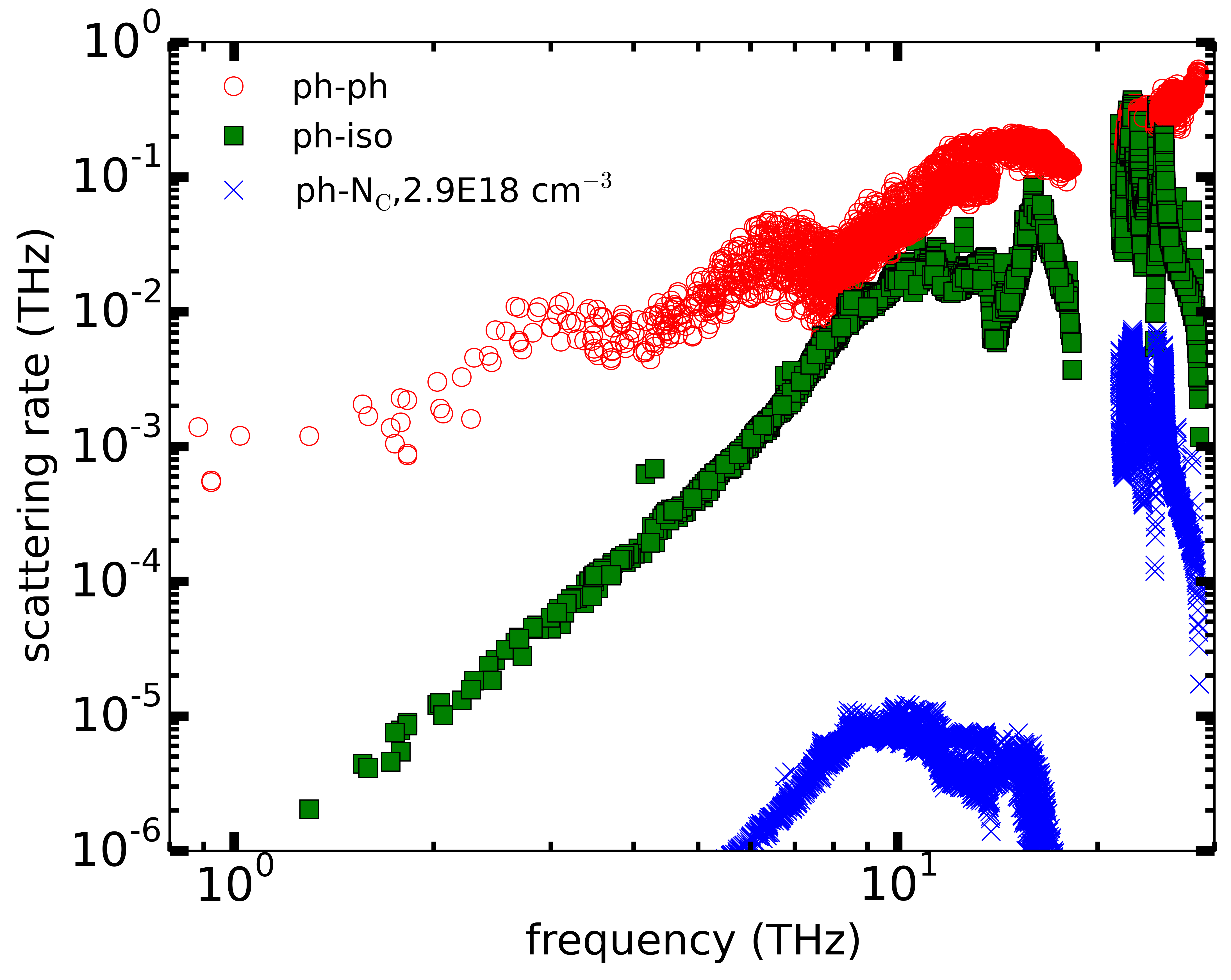}
    \caption{Phonon-phonon, phonon-isotope and phonon-N-substitution scattering rates for 6H SiC at 300 K. N substitution is considered at the C site with a concentration of $2.9\times10^{18}$ \pcm $ $ matching the highest N-doped sample in Ref. \citep{morelli1993eph}.}
    \label{fig:scatrate}
\end{figure}

In the 4H and 6H SiC samples considered here, substitutional nitrogen defects are present, and these can also be treated using the mass disorder model. We do not include bond-disorder arising from these substitutional defects. Here we assume that the nitrogen defects are at the carbon sites. In Fig. \ref{fig:scatrate} the phonon-phonon, phonon-isotope and phonon-nitrogen defect scattering rates are shown for the 6H phase at $300$ K. Phonon-phonon scattering dominates over the entire frequency range, followed by phonon-isotope scattering. The sample with the highest impurity concentration to which we compare our calculations has $2.9\times10^{18}$ cm$^{-3}$ N-atoms. Treating these as substitutional defects at the C site leads to negligible reduction in $\kappa$. This is because N defects give only a small ($17 \%$) mass increase over the C atoms, only slightly larger than that for the minority carbon isotope ($^{13}$C), and the N concentration is $300$ times smaller than that of $^{13}$C in naturally occurring carbon. The low concentration of N defects means that even if the N atoms were on the Si site instead of  the C site, the phonon defect scattering rates would still be negligibly small. To support this, we note that in a recent study on cubic 3C SiC \citep{katre2017exceptionally}, N substitution defects at the C site were treated as both mass and bond defects. For the N-doping concentrations considered in the present work, a negligible reduction in the $\kappa$ of 3C SiC was found around $300$ K. For all $\kappa$ calculations phonon-isotope scattering is included, while phonon-substitution defect scattering is ignored for the remaining results presented.

\begin{figure}
    \centering
    \includegraphics[width=1.0\linewidth]{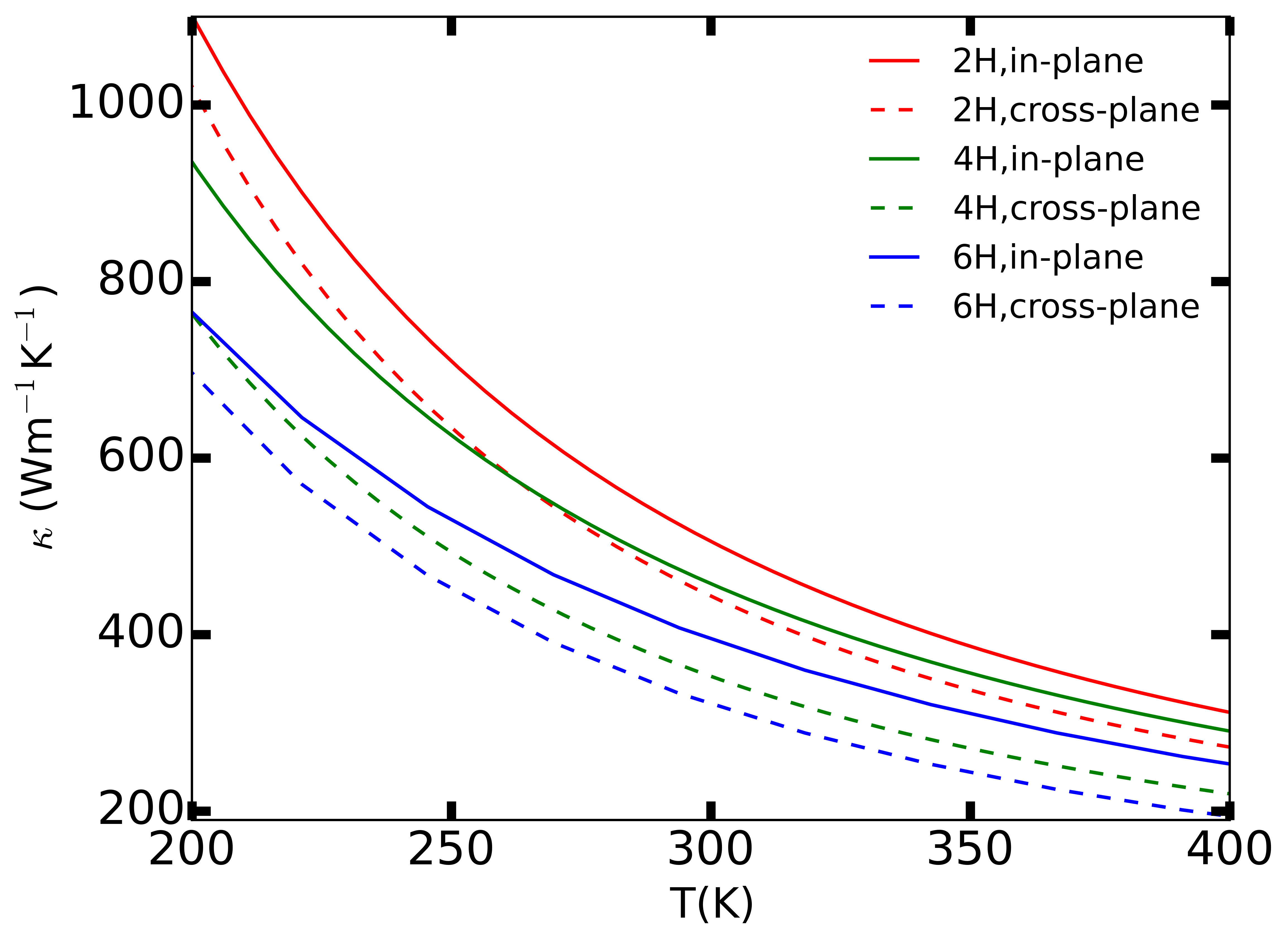}
    \caption{Temperature dependence of $\kin$ and $\kout$ for 2H, 4H and 6H SiC. For a given direction, $\kappa_{2H} > \kappa_{4H} > \kappa_{6H}$. For both the solid and dashed types, the top (middle) [bottom] line corresponds to 2H (4H) [6H] SiC $\kappa$.}
    \label{fig:kappa}
\end{figure}

Fig. \ref{fig:kappa} shows the calculated temperature dependence of $\kin$ and $\kout$ for the three polytypes considered. For both in-plane and cross-plane directions, we find that $\kappa_{\text{2H}} > \kappa_{\text{4H}} > \kappa_{\text{6H}}$. This is explained by the fact that in these hexagonal polytypes the larger number of atoms with increasing $n$ in $n$H SiC gives rise to an increasing number of low-lying optic branches. The small phonon group velocities in these branches combined with the fact that the three-phonon scattering rates are the highest in 6H SiC and the lowest in 2H SiC leads to lower $\kappa$ with increasing $n$ in $n$H SiC. Also, for given $n$ in $n$H SiC, we find that $\kin > \kout$ over the temperature range considered.  This anisotropy arises from larger contributions to the in-plane thermal conductivity integral, Eq. \ref{eq:kappa}, from the $v_{\lambda}^{x}F_{\lambda}^{x}$ terms compared to those for the corresponding cross-plane ($z$) components, especially in the frequency range around $10$ THz. Roughly the same anisotropy is found in both the RTA and the iterated calculations.

\begin{figure}[!ht]
    \centering    
    \subfloat[]{%
      \includegraphics[width=1.0\linewidth]{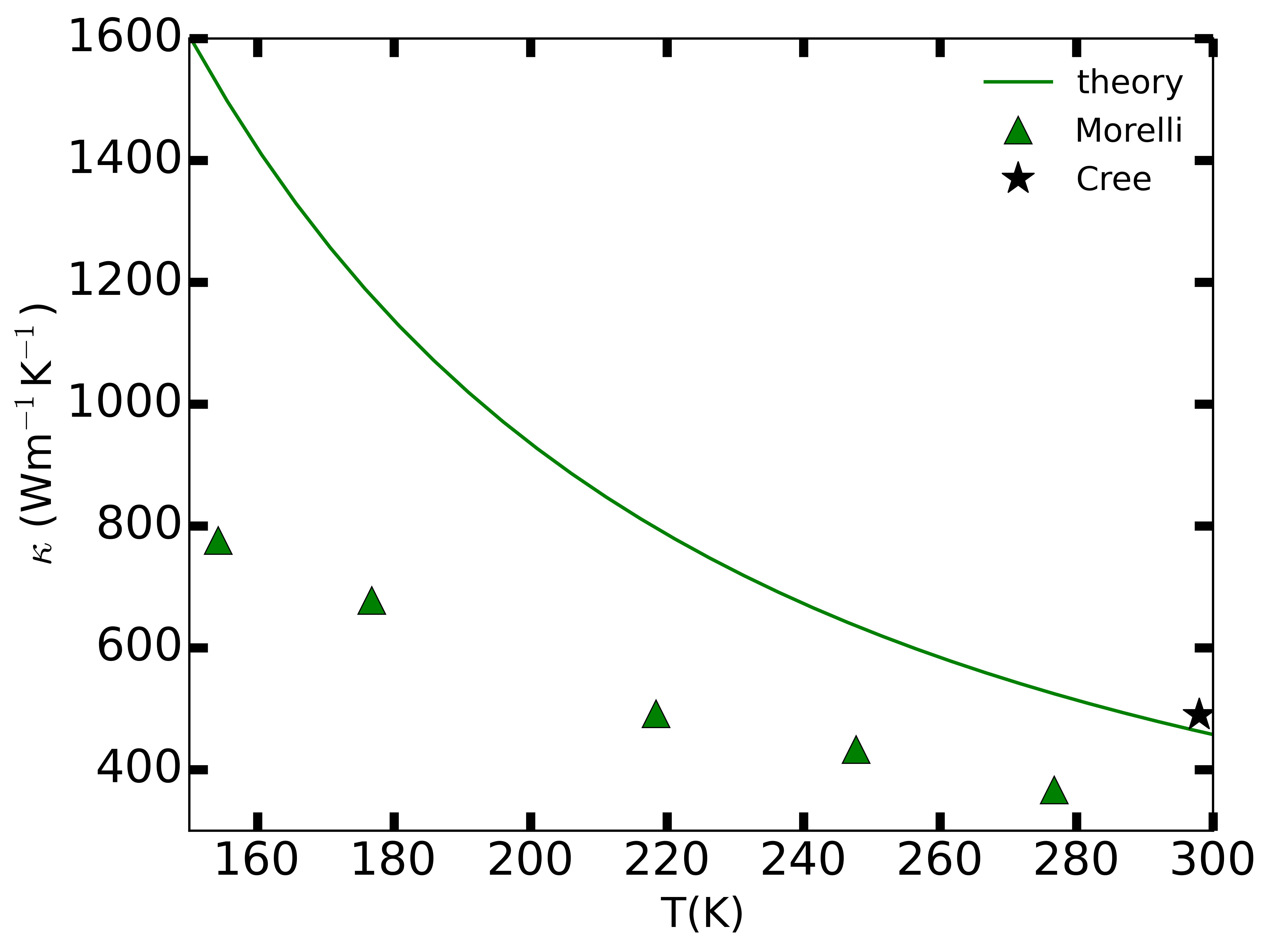}
    }\\
    \subfloat[]{%
      \includegraphics[width=1.0\linewidth]{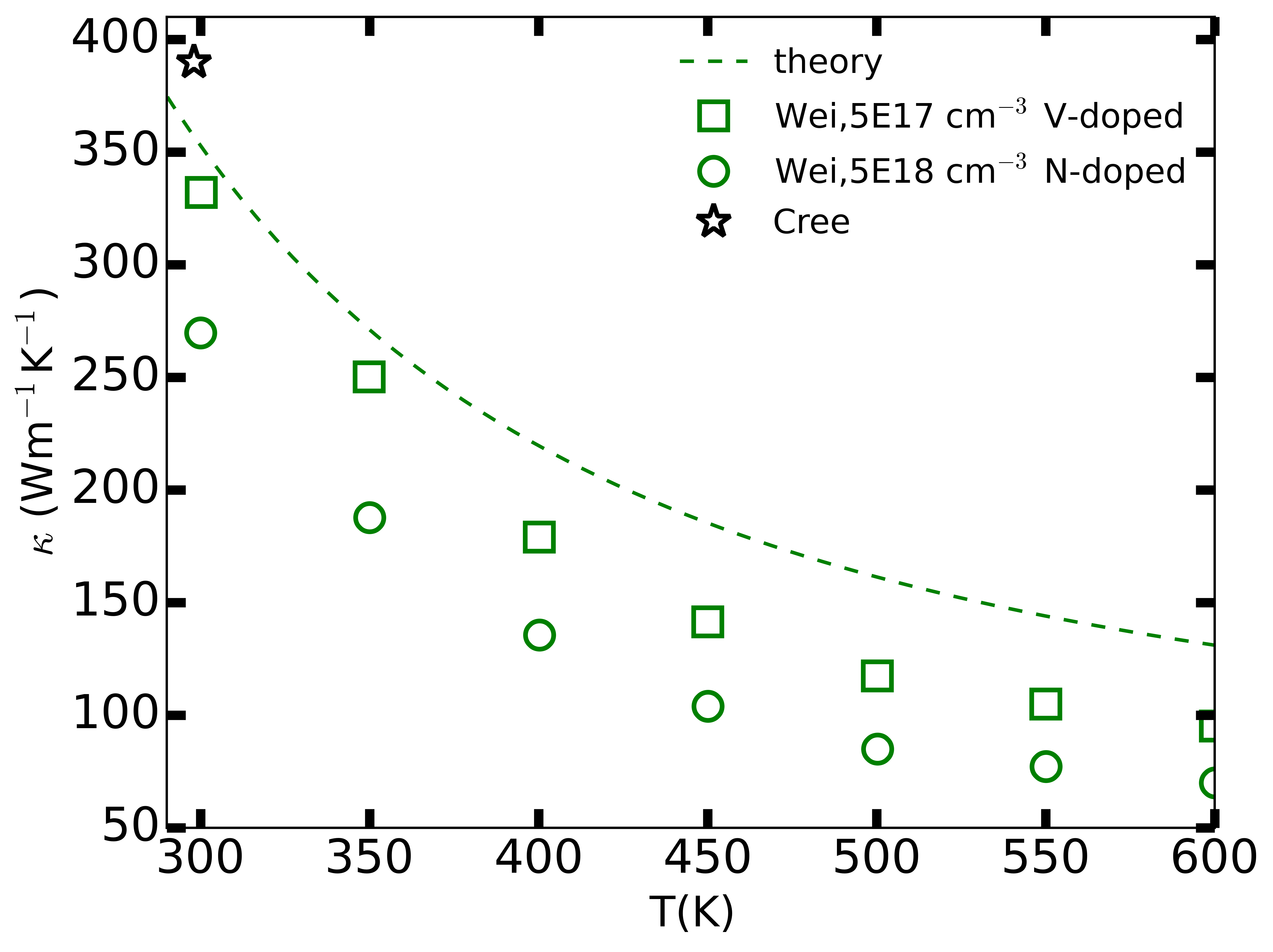}
    }
    \caption{Comparison of the calculated 4H SiC (a) $\kin$ and (b) $\kout$ with experimental measurements from \citep{morelli1993conf} (green triangles), \citep{wei2013thermal} (green open squares and circles) and \citep{cree} (black open stars).}
    \label{fig:4hk}
\end{figure}

In Fig. \ref{fig:4hk} (a) the calculated $\kin$ of 4H SiC are compared to experimental data. Green triangles give $\kin$ measured by Morelli et al. \citep{morelli1993conf}. These data points lie well below the calculated $\kin$ values (solid green curve). The authors in Ref. \citep{morelli1993conf} speculate that their 4H sample may have stacking faults. These types of defects are not treated in our theoretical framework, and we attribute the large difference between calculation and the data from Ref. \citep{morelli1993conf} to the presence of such defects. On the other hand, a measurement of $\kin$ by Cree Inc. \citep{cree} (black filled symbol) on their high purity semi-insulating sample at 298 K shows good agreement with the calculated $\kin$. Specifically, we calculated $\kin = 451$ Wm$^{-1}$K$^{-1}$, only $8 \%$ smaller than the measured Cree value of $490$ Wm$^{-1}$K$^{-1}$. 

Fig. \ref{fig:4hk} (b) shows the calculated results for $\kout$ (green dashed curve) compared to measured data from Wei et al. \citep{wei2013thermal} and that from Cree \citep{cree}. The sample with higher $\kout$ from Ref. \citep{wei2013thermal} was believed to contain $5\times 10^{17}$ \pcm $ $ vanadium substitutional defects while that with the lower $\kout$ contained $5\times 10^{18}$ \pcm $ $ nitrogen substitutional defects. Treating the V and N dopants as substitutional defects at the C site lead to negligible reduction of $\kappa$. The measurements near room temperature from the less doped samples of Ref. \citep{wei2013thermal} and from Ref. \citep{cree} are in good agreement with the calculated results. However, at higher temperatures, both sets of measured data from Ref. \citep{wei2013thermal} lie well below our calculated values. This is puzzling to us. Moreover, the measured $\kappa$ values for the sample with the higher defect/carrier density are significantly smaller than those for the lower defect/carrier density sample even though the calculated phonon-defect scattering rates in both cases are small and phonon-electron scattering in  this temperature range is weak (see discussion below).  Finally, the measured $\kappa$ values for both  samples saturate above $600$ K, a finding inconsistent with the roughly 1/$T$ dependence of the thermal conductivity from phonon-phonon scattering and the expected negligible contribution from the electronic part of the thermal conductivity.

\begin{figure}
    \centering
    \includegraphics[width=1.0\linewidth]{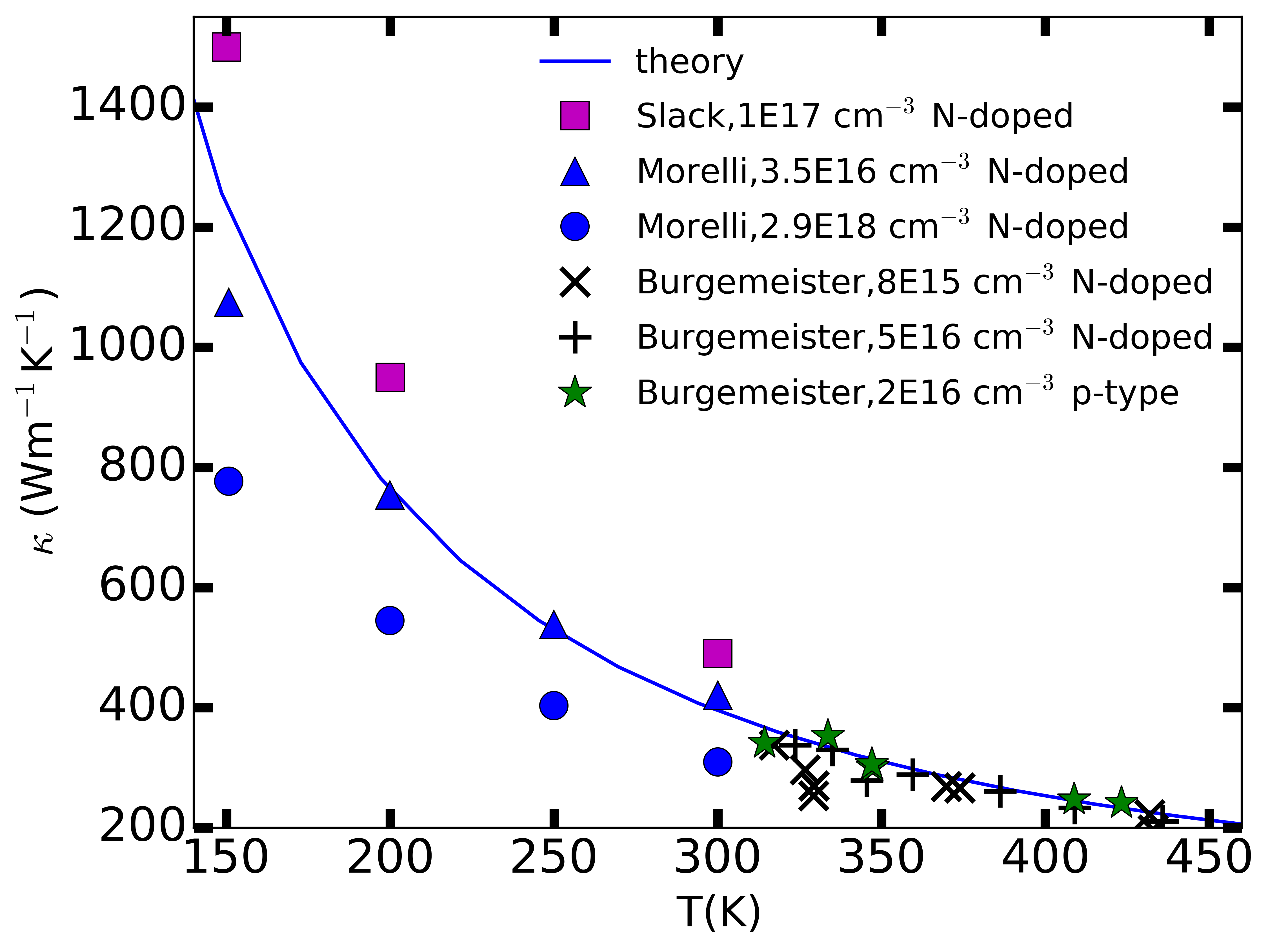}
    \caption{Comparison of the calculated 6H SiC $\kin$ with measured data from Ref. \citep{slack1964thermal} (purple squares), Ref. \citep{morelli1993eph} (blue triangles and circles) and Ref. \citep{burgemeister1979thermal} (green stars, black crosses and black pluses).}
    \label{fig:6hkin}
\end{figure}

In Fig. \ref{fig:6hkin} we compare $\kin$ of 6H SiC with measurements from Slack \citep{slack1964thermal}, Morelli et al. \citep{morelli1993eph} and Burgemeister et al. \citep{burgemeister1979thermal}. Calculated values (solid blue line) lie below the Slack data (purple squares) with N defect concentration of $1\times10^{17}$ \pcm, but above the data from Morelli et al. whose samples had N concentrations of $3.5\times10^{16}$ and $2.9\times10^{18}$ \pcm, respectively. For the sparsely doped Morelli sample (blue triangles) agreement with theory is very good between $200$ and $300$ K - at $150$ K the calculated value is about $20 \%$ larger than the experimental one. For the higher doped Morelli sample (blue circles) the measured $\kin$ is significantly lower than the calculated values. This reduction is attributed by Morelli et al. to phonons scattering from electrons donated by N atoms. The Burgemeister data points (black symbols) are measurements above $300$ K for lightly N- or p-type doped samples. There is some variation in the measured data, but in general the trend is well captured by the theory curve. We note that the $8\times10^{15}$ \pcm $ $ N-doped Burgemeister data points (black cross symbols) slightly above $300$ K show a variation nearly equal to the difference between the nearest Morelli data points. Moreover, the Slack sample with nearly $3$ times the N-doping of the the purest Morelli sample has a significantly higher $\kappa$ than the latter and also projects to larger $\kin$ than the higher purity samples of Burgemeister. The reason for the high measured $\kin$ for the Slack sample is unclear to us. It is also unclear whether the difference between the $\kappa$ measurement on the two Morelli samples is entirely due to the EPI.

\begin{figure}
    \centering
    \includegraphics[width=1.0\linewidth]{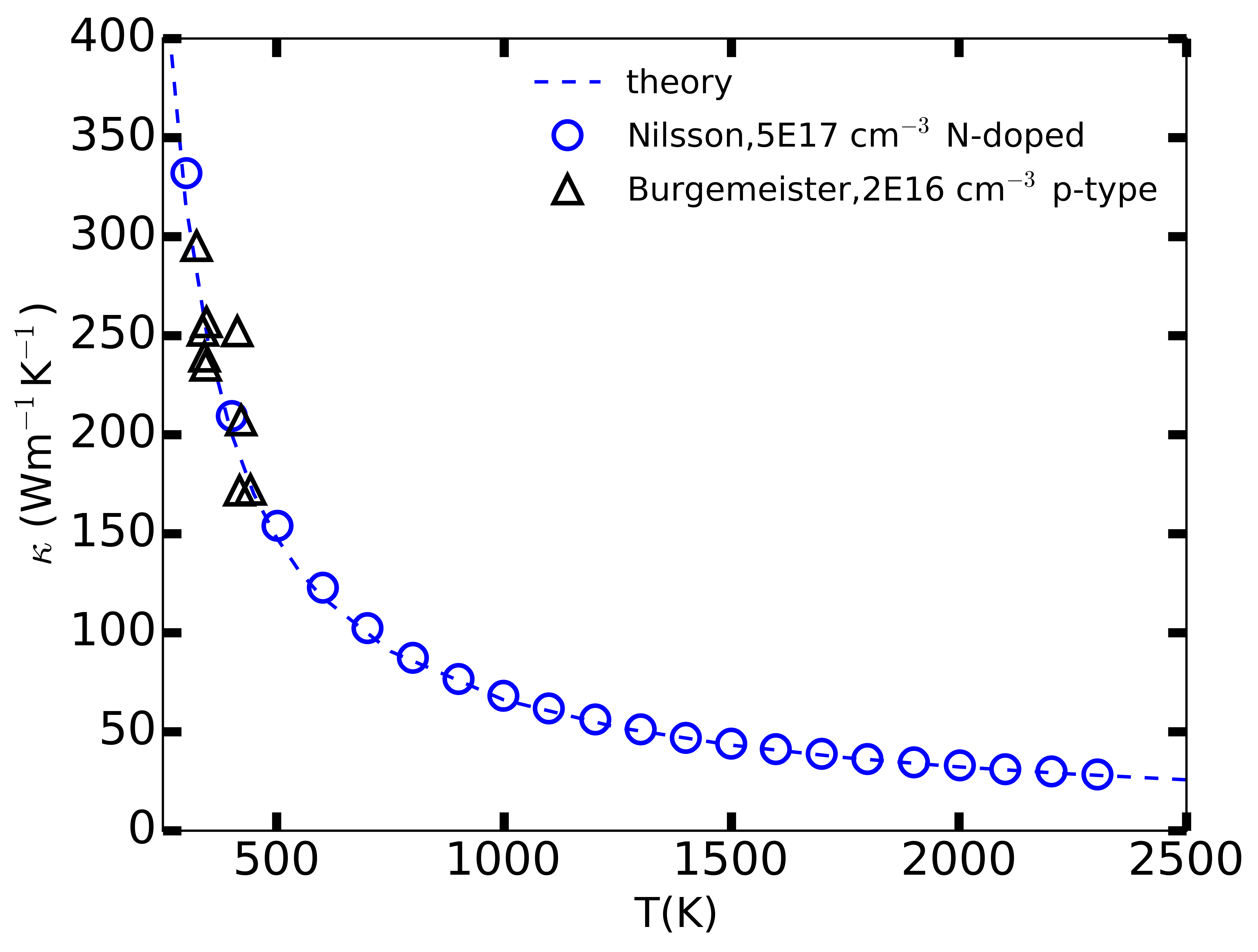}
    \caption{Comparison of 6H SiC $\kout$ with experimental measurements from Refs. \citep{nilsson1997determination} (blue open circles) and \citep{burgemeister1979thermal} (black open triangles).}
    \label{fig:6hkout}
\end{figure}

In Fig. \ref{fig:6hkout} the calculated $\kout$ of 6H SiC is compared to experimental data from Nilsson et al. \citep{nilsson1997determination} between $300$ and $2300$ K. Also plotted are $\kout$ measurements by Burgemeister et al. \citep{burgemeister1979thermal}. Excellent agreement between calculation and data from Refs. \citep{nilsson1997determination} and \citep{burgemeister1979thermal} is found in this case over the entire temperature range of the data. We note that even at high temperatures the thermal expansion coefficient of 6H SiC is almost independent of temperature \citep{taylor1960silicon}, which suggests that higher-order anharmonicity is weak even at the high temperatures (up to 2300K) of the measurements of Ref. \citep{nilsson1997determination}. 

The experimental sample from Ref. \citep{nilsson1997determination} has an N-doping concentration of $5\times10^{17}$ \pcm. At $300$ K the measured $\kappa$ value is $4.1 \%$ higher than the calculated value. Interestingly, while the EPI was thought to give a significant reduction in $\kin$ in Ref. \citep{morelli1993eph}, \citep{burgemeister1979thermal}, \cite{morelli1993conf} even at 300K, such a reduction is not seen by comparing the $300$ K $\kout$ data from Nilsson and Burgemeister in spite of there being an order of magnitude difference in carrier densities. Instead the two sets of $\kout$ measured values are quite close to each other and are also in very good agreement with the theoretical results.

\begin{figure}[!ht]
    \centering
    \includegraphics[width=1.0\linewidth]{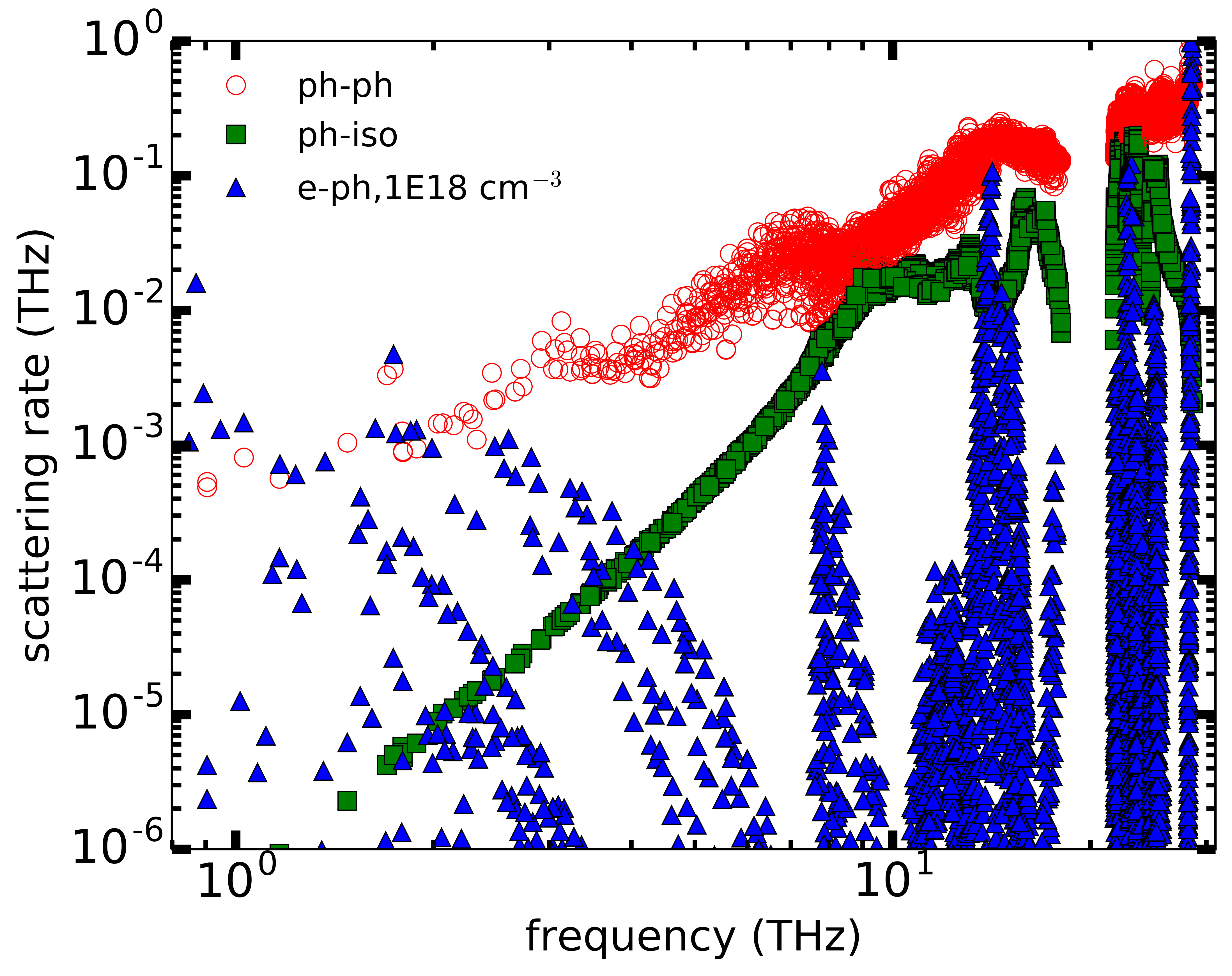}
    \caption{Phonon-electron scattering rates in 2H SiC at $300$ K for an electron concentration of $10^{18}$ \pcm. For comparison, phonon-phonon and phonon-isotope scattering rates are also plotted.}
    \label{fig:2heph}
\end{figure}

\begin{figure}[!ht]
    \centering
    \includegraphics[width=1.0\linewidth]{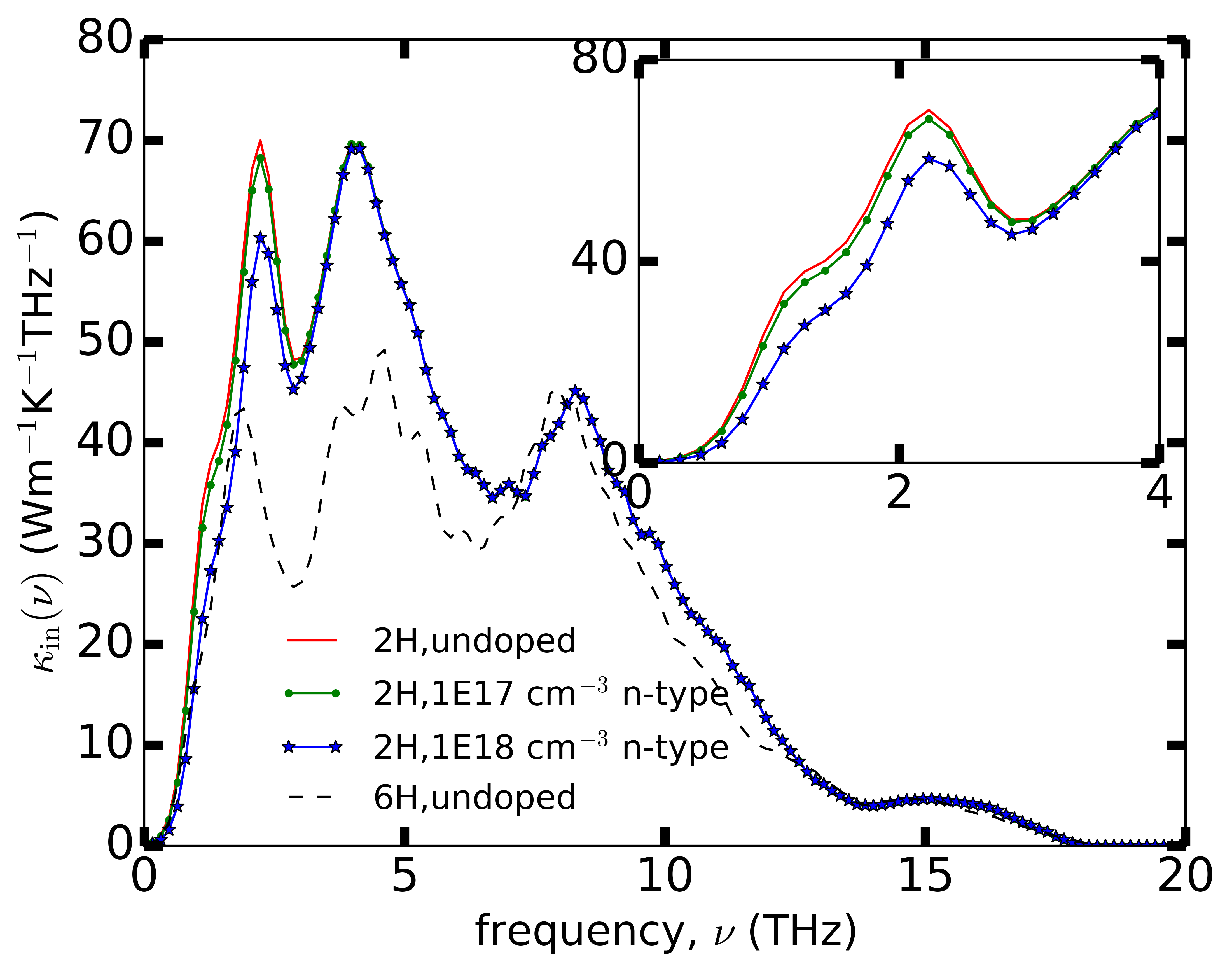}
    \caption{Spectrum of $\kin$ as a function of frequency $\nu$ for 2H and 6H SiC at $300$ K. Effect of EPI is shown for 2H for two n-type doping levels. The inset plot zooms in on the [$0$,$4$] THz region.}
    \label{fig:kinspec}
\end{figure}

To address the question of to what extent the EPI reduces $\kappa$ we calculated phonon-electron scattering rates for the 2H phase. Full first principles treatment of the EPI in 4H and 6H SiC is currently beyond our computational capabilities. In Fig. \ref{fig:2heph} phonon-electron scattering rates for 2H SiC for a representative electron density of $10^{18}$ \pcm $ $ are plotted along with phonon-phonon and phonon-isotope scattering rates at $300$ K. The phonon-electron scattering rates are shown here on a dense $q$-mesh of $50\times50\times30$ to highlight their frequency dependence while the other scattering rates are shown on the originally converged $25\times25\times15$ $q$-mesh. In the low frequency region the acoustic-phonon-electron scattering rates are comparable to the phonon-phonon scattering rates before quickly falling off above $2$ THz. The phonon-electron scattering rates become strong again for the high optic phonon frequencies. The high-lying optic phonons contribute negligibly to $\kappa$, as can be seen from the spectral contribution to $\kin$ ($\kin(\nu)$) \citep{knu} for 2H and 6H SiC shown in Fig. \ref{fig:kinspec}. Therefore, strong phonon-electron scattering for those phonons will not affect $\kappa$. It is seen in Fig. \ref{fig:kinspec} that the effect of phonon-electron scattering in suppressing the $\kin$ of 2H SiC is only visible below around $3$ THz.  For example, at $1$ THz, $\kin(\nu)$ is suppressed by about $39 \%$, but beyond $3$ THz there is no suppression. At higher temperatures above $300$ K phonon-phonon scattering dominates over all other scattering mechanisms and the effect of the EPI is predicted here to be negligible. However, at lower temperatures, phonon-phonon scattering weakens and the reduction in $\kappa$ due to phonon-electron scattering can become significant.

The effect of the EPI on 2H SiC $\kin$ is presented in Table \ref{tab:2hepi} for $10^{17}$ \pcm $ $ and $10^{18}$ \pcm $ $ n-type doping concentrations. For $10^{17}$ \pcm $ $ doping concentration we see a $20 \%$ reduction of $\kin$ at $100$ K while the reductions at $300$ K and $600$ K are negligible. For $10^{18}$ \pcm $ $ doping concentration these numbers are $28 \%$, $4.4 \%$ and $1.1 \%$, respectively. These results are consistent with the discussion in the previous paragraph.

Table \ref{tab:6hepi} compares the calculated undoped $\kin$ for 6H SiC with the corresponding measured values from Ref. \citep{morelli1993eph} for the sample with electron density of $2.9\times10^{18}$ \pcm. Assuming that the strength of the EPI in 6H SiC is similar to that in 2H SiC and coupling this to the fact that phonon-phonon scattering is stronger in 6H SiC than in 2H SiC we conclude that the EPI is not likely the only reason for the strong reduction in $\kappa$ in Ref. \citep{morelli1993eph}, nor can it be the sole explanation for the difference in thermal conductivities of the 6H samples from Ref. \citep{slack1964thermal} (see Fig. \ref{fig:6hkin}) or the 4H samples from Ref. \citep{wei2013thermal} (see Fig. \ref{fig:4hk}b).

\begin{table}
    \centering
    \begin{tabular}{c|c|c|c}
        T(K) & undoped & n-type & n-type \\
             &          & $10^{17}$ \pcm & $10^{18}$ \pcm \\
        \hline
        $100$ & $4741$ & $3779$, $(20 \%)$  & $3398$, $(28 \%)$\\
        $300$ & $497$ & $492$, $(1.0 \%)$ & $475$, $(4.4 \%)$\\
        $600$ & $186$ & $185$, $(0.5 \%)$ & $184$, $(1.1 \%)$\\
    \end{tabular}
    \caption{Calculated 2H SiC $\kin$ (Wm$^{-1}$K$^{-1}$) with and without the EPI. The third and fourth columns give $\kin$, percentage reduction with respect to the undoped case.}
    \label{tab:2hepi}
\end{table}

\begin{table}
    \centering
    \begin{tabular}{c|c|c}
        T(K) & undoped & N-doped \\
             & (theory) & $2.9\times10^{18}$ \pcm \\
        \hline
        $100$ & $2460$ & $1018$, $(59 \%)$\\
        $300$ & $382$ & $310$, $(19 \%)$\\
    \end{tabular}
    \caption{6H SiC $\kin$ (Wm$^{-1}$K$^{-1}$) values from theory (undoped, column 2) .vs. experiment (N-doped, column 3) \citep{morelli1993eph}. The third column also shows percentage difference between the calculated and measured $\kin$.}
    \label{tab:6hepi}
\end{table}

\section{Conclusion} \label{conclusion}

We have presented \textit{ab initio} calculations of the in-plane and cross-plane lattice thermal conductivities, $\kin$ and $\kout$, for 2H, 4H and 6H SiC. We found that for both in-plane and cross-plane thermal transport, the 2H phase has the highest $\kappa$, followed by that of 4H SiC and then by that of 6H SiC. Also, for given $n$ in $n$H SiC, $\kin$ is found to be larger than $\kout$ over a large range of temperatures. Reasonably good agreement is obtained with the measured $\kin$ and $\kout$ results for some of the 4H and 6H samples. However, wide variations in measured data were noted that in most cases suggest the presence of defects in the measured samples. By calculating phonon-electron scattering from first principles for the 2H phase at electron densities representative of those in the studied samples, we concluded that such scattering is not likely the sole reason for the difference in $\kappa$ measurements between the 6H SiC samples considered in Ref. \citep{morelli1993eph} or for the difference between the measured $\kappa$ values for the 4H SiC samples in Ref. \citep{wei2013thermal}. Nevertheless, carrier concentrations as low as $1\times10^{18}$ \pcm $ $ in the 2H phase can lead to a significant reduction in $\kappa$ at low temperatures, which is qualitatively consistent with the measurement findings \citep{morelli1993eph}.

\section*{Acknowledgements}

N.H.P. and D.B. acknowledge support from the Office  of Naval Research MURI, Grant No. N00014-16-1-2436 and from the Pleiades computational cluster of Boston College. This work used the Extreme Science and Engineering Discovery Environment (XSEDE) \citep{towns2014xsede}, which is supported by National Science Foundation grant number ACI-1548562. N.H.P. acknowledges help from Dr. Chunhua Li with \texttt{EPW} calculations. L.L. acknowledges support from the U.S. Department of Energy, Office of Science, Office of Basic Energy Sciences, Materials Sciences and Engineering Division. AK and NM acknowledge support from the Air Force Office of Scientific Research, USAF under award No. FA9550615-1-0187 DEF.

\appendix

\section{}
In this appendix we provide the expressions for the phonon-phonon and phonon-mass defect scattering rates.

Three phonon scattering processes satisfy the following quasimomentum and energy conservation relations
\begin{align}\label{eq:procs}
    &\mathbf{q} \pm \mathbf{q}' + \mathbf{G} = \mathbf{q}'' \nonumber \\
    &\omega_{\lambda} \pm \omega_{\lambda'} = \omega_{\lambda''},
\end{align}
where $\mathbf{G}$ a the reciprocal lattice vector.

For the $\pm$ processes defined in \ref{eq:procs} the three phonon scattering matrix elements are given by
\begin{equation}
    \Phi^{\pm}_{\lambda\lambda'\lambda''} = \sum_{<i>jk}\sum_{\alpha\beta\gamma}\Psi^{\alpha\beta\gamma}_{ijk}\dfrac{e^{i\alpha}_{s,\mathbf{q}}e^{j\beta}_{s',\pm\mathbf{q}'}e^{k\gamma}_{s'',-\mathbf{q}''}}{\sqrt{m_{i}m_{j}m_{k}}},
\end{equation}
where $i,j,k$ label atoms in the supercell with $<>$ symbolizing restricted sum over the primitive cell, $m_{i}$ is the mass of atom $i$, $e^{i\alpha}_{s,\mathbf{q}}$ is the $\alpha$ Cartesian component of the phonon eigenvector associated with polarization $s$ and wavevector $\mathbf{q}$, and $\Psi^{\alpha\beta\gamma}_{ijk} = \dfrac{\partial^{3}U}{\partial r^{\alpha}_{i}\partial r^{\beta}_{j}\partial r^{\gamma}_{k}}$ is the third-order anharmonic force constant where $r^{\alpha}_{i}$ is the displacement of atom $i$ in the Cartesian direction $\alpha$ calculated from the crystal potential energy $U$.

In terms of the scattering matrix elements the three phonon scattering rates of single processes are
\begin{align} 
    W^{+}_{\lambda\lambda'\lambda''} &= \dfrac{\hbar\pi}{4}\dfrac{n^{0}_{\lambda'} - n^{0}_{\lambda''}}{\omega_{\lambda}\omega_{\lambda'}\omega_{\lambda''}} |\Phi^{+}_{\lambda\lambda'\lambda''}|^{2} \nonumber \\
    &\times\delta(\omega_{\lambda} + \omega_{\lambda'} - \omega_{\lambda''}), \label{eq:phphplus} \\  
    W^{-}_{\lambda\lambda'\lambda''} &= \dfrac{\hbar\pi}{4}\dfrac{n^{0}_{\lambda'} + n^{0}_{\lambda''} + 1}{\omega_{\lambda}\omega_{\lambda'}\omega_{\lambda''}} |\Phi^{-}_{\lambda\lambda'\lambda''}|^{2} \nonumber \\  
    &\times \delta(\omega_{\lambda} - \omega_{\lambda'} - \omega_{\lambda''}). \label{eq:phphminus}
\end{align}

Phonon-mass defect scattering rates for single processes are given by
\begin{equation} \label{eq:phiso}
    W_{\lambda\lambda'} = \dfrac{\pi}{2}\omega^{2}_{\lambda}\sum_{i}g_{i}|\mathbf{e}^{*}_{i\lambda}\cdot\mathbf{e}_{i\lambda'}|^{2}\delta(\omega_{\lambda}-\omega_{\lambda'}),
\end{equation}
where $i$ labels an atom in the primitive cell and
\begin{equation}\label{eq:massvar}
    g_{i} = \sum_{t}f_{ti}\left(\dfrac{\Delta m_{ti}}{\overline{m}_{i}}\right)^{2}
\end{equation}
is the mass variance parameter for atom $i$, with $\Delta m_{ti} \equiv m_{ti} - \overline{m}_{i}$ and $t$ denoting the type of mass defect (eg. isotope, substitutional defect) at site $i$. $\overline{m}_{i}$ is the average mass of the atom $i$ and $f_{it}$, the fraction of atoms of type $t$ at the site $i$.

In the RTA the total phonon scattering rates including phonon-phonon and phonon-mass defect scattering are
\begin{align} \label{eq:tau0tot}
    &\left(\tau^{\text{anh}}_{\lambda}\right)^{-1} = \dfrac{1}{N}\bigg(\sum_{\lambda'\lambda''}^{+}W^{+}_{\lambda\lambda'\lambda''} + \dfrac{1}{2}\sum_{\lambda'\lambda''}^{-}W^{-}_{\lambda\lambda'\lambda''}\bigg) \\ 
    &\left(\tau^{\text{mass}}_{\lambda}\right)^{-1} = \dfrac{1}{N}\sum_{\lambda'}W_{\lambda\lambda'},
\end{align}
where $N$ is the total number of points in the $\Gamma$-centered, regular $q$-mesh sampling the Brillouin zone and the $\pm$ on the summation symbolize conservation restrictions given by \ref{eq:procs}. And finally, the function $\mathbf{\Delta}$ appearing in the linearized PBE is given by

\begin{align}\label{eq:delta}
    \mathbf{\Delta}_{\lambda} &= \dfrac{1}{N\omega_{\lambda}}\bigg\{\sum^{+}_{\lambda'\lambda''} W^{+}_{\lambda\lambda'\lambda''}\left(\omega_{\lambda''}\mathbf{F}_{\lambda''} - \omega_{\lambda'}\mathbf{F}_{\lambda'}\right) \nonumber \\
    &+ \dfrac{1}{2}\sum^{-}_{\lambda'\lambda''} W^{-}_{\lambda\lambda'\lambda''}\left(\omega_{\lambda''}\mathbf{F}_{\lambda''} + \omega_{\lambda'}\mathbf{F}_{\lambda'}\right)\bigg\}. 
\end{align}

The energy conserving delta functions appearing in equations above are approximated in \texttt{ShengBTE} using a Gaussian function
\begin{equation} \label{eq:gauss}
    \delta(\omega_{\lambda} - \omega) \approx \dfrac{1}{\sqrt{2\pi}\sigma_{\omega}}\text{exp}\left[-\dfrac{(\omega_{\lambda} - \omega)^{2}}{2\sigma_{\omega}^{2}}\right],
\end{equation}
where $\sigma_{\omega}$ is a locally adaptive broadening parameter found using
\begin{equation}\label{eq:adaptive}
    \sigma_{\omega} \approx \dfrac{1}{\sqrt{12}}\sqrt{\sum_{\mu}\left[(\mathbf{v}_{\lambda'}-\mathbf{v}_{\lambda''})\cdot\dfrac{\mathbf{Q}_{\mu}}{N_{\mu}}\right]^{2}}, 
\end{equation}
where $\mathbf{Q}_{\mu}$ is the reciprocal lattice vector in the direction $\mu$, $\mathbf{v}_{\lambda}$ is the Cartesian group velocity of the phonon mode $\lambda$, and $N_{\mu}$ is the number of grid points used along the reciprocal lattice vector $\mathbf{Q}_{\mu}$. In $n$H SiC, $Q_{3} < Q_{1(2)}$. If the same $N_{\mu} = N$ is used for all three directions then $Q_{3}/N < Q_{1(2)}/N$. As a result $\Gamma$-A phonon modes, for which $\mathbf{q} = (0,0,q_{z})$, will have selectively smaller $\sigma_{\omega}$ compared to other modes. This is why $N_{\mu}$ must be chosen such that the grid spacing along all three directions is uniform.


\bibliographystyle{elsarticle-num}

\bibliography{refs}

\begin{thebibliography}{10}
\bibitem{2hdisp}
To our knowledge, there is no measured phonon data for 2{H} {S}i{C}.
\bibitem{knu}
The spectrum of $\kappa^{\alpha\beta}$ is given by $\kappa^{\alpha\beta}(\nu) =
  \dfrac{k_{\text{B}}}{NV}\sum_{\lambda}\left(\dfrac{\hbar
  \omega_{\lambda}}{\text{k}_{B}T}\right)^{2}n^{0}_{\lambda}(n^{0}_{\lambda} +
  1)v^{\alpha}_{\lambda}F^{\beta}_{\lambda}\delta(\nu - \nu_{\lambda})$, where
  $(2\pi)^{-1}\nu$ is the frequency.
\end{thebibliography}

\end{document}